# Unlocking the Potential of GeS Monolayer: Strain-Enabled Control of Electronic Transports and Exciton Radiative Lifetimes

Vo Khuong Dien[1], Pham Thi Bich Thao[2], Nguyen Thi Han[1], Nguyen Duy Khanh[3], Le Vo Phuong Thuan[1], Ming-Fa Lin[1,4], and Nguyen Thanh Tien[2].

[1]Department of Physics, National Cheng Kung University, 701 Tainan, Taiwan
[2]College of Natural Sciences, Can Tho University, 3-2 Road, Can Tho City 94000, Vietnam
[3]High-Performance Computing Laboratory (HPC Lab), Thu Dau Mot University, 75100, Binh Duong, Vietnam
[4]Hierarchical Green-Energy Material (Hi-GEM) Research Center, National Cheng Kung University, 701 Tainan, Taiwan
Email: vokhuongdien@gmail.com, and nttien@ctu.edu.vn

Monolayer germanium sulfide (GeS) is gaining significant attention for its exceptional anisotropic electronic conductance, notable excitonic effects, and wide range of potential applications. In our study, we used density functional theory (DFT), many-body perturbation theory (MBPT), and non-equilibrium Green's function (NEGF) to investigate electronic transport properties and exciton radiative lifetime of single-layer germanium sulfide. Our theoretical findings showed that applying up to 8% compressive strain increased carrier mobility by nearly threefold, and thus, dramatically enhance the device's current intensity. Moreover, we observed that strain engineering allowed fine-tuning of the electron-hole recombination time. At 6% tensile strain, the effective radiative lifetime was as short as 19 picoseconds, which is 4.5 times faster than the intrinsic state and 80 times faster than at 8% compressive strain. These results highlight the potential of strain engineering to customize the electronic and optical properties of GeS monolayer for specific electronic, optoelectronic, and photovoltaic device requirements.

## 1. Introduction

The successful fabrication of monolayer graphene [1] has sparked significant interest in exploring other two-dimensional (2D) materials, including transition metal chalcogenides (TMDCs) [2], group III mono-chalcogenides [3, 4], hexagonal boron nitride (h-BN) [5, 6], phosphorene [7], and others. Recently, 2D germanium sulfide (GeS) has emerged as a highly researched material [8-10]. Like black phosphorus, bulk GeS also adopts a layered structure with weak Van-der-Waals (vdWs) interactions between interlayers and strong covalent bonding within layers. The few layers of GeS have been successfully fabricated via either vapor transport process [9] or mechanical exfoliation [10], while the monolayer GeS is predicted to be dynamic stability suggesting the high ability to exfoliate the monolayer GeS from its bulk counterpart. In contrast to semimetal graphene, monolayer GeS has a sizable electronic band gap (~2.3 eV) [11] making it suitable for semiconductor applications. Additionally, the monolayer form of GeS is predicted to have a larger free carrier mobility (~$10^3$ $cm^2.V^{-1}.s^{-1}$) [11] compared to $MoS_2$ (~200 $cm^2.V^{-1}.s^{-1}$) [12]. As a result of the ultrathin monolayer and the significant reduction of dielectric screening, the excitonic effects are predicted to be very strong in the GeS single layer [13, 14]. One notable feature of GeS is its anisotropic electric conductance and optical responses [15], which distinguish it from isotropic 2D crystals such as graphene and MoS$_2$. Therefore, it would be exciting to explore ways to manipulate these anisotropies further.

The current research focus in the field of condensed matter physics involves modifying the electronic and optical properties of layered materials [16, 17]. This can be achieved through the introduction of ad-atoms [18, 19], the application of electric and magnetic fields [20], the adsorption of molecule clusters [16], and the creation of defects [21, 22]. Another effective method for altering the properties of materials is strain engineering, which is particularly useful for one-dimensional [23] and two-dimensional [24] crystals due to their ability to withstand larger strains compared to bulk crystals. For instance, monolayer graphene [25] and MoS$_2$ [26] can sustain strains up to their intrinsic limit (approximately 15% for graphene and 11% for MoS$_2$) without causing significant damage to their crystal structures. This provides a wide range of opportunities for tuning their mechanical and electronic performances.

Herein, by combining density functional theory (DFT) [27], the many-body perturbation theory (MBPT) [28], and the non-equilibrium Green's Function (NEGF) [29], we illustrated that strain engineering can serve as an effective tool to tailor the electronic transport properties and the recombination time scale of exciton states. Our theoretical calculations and analytical analyses indicated that electron mobility could be significantly enhanced under compressive strain, the current-voltage (I-V) characteristic of the device, additionally, shows an extremely high current intensity. Moreover, the excitonic effects, especially the exciton radiative lifetime can be fine-tuned upon applying the external strains. The theoretical results achieved in the current research are paramount not only for basic sciences but also for high-tech applications, such as ultrafast field-effects transistors (FETs), light-emitting-diodes (LEDs), and photovoltaic (PVs) applications.

## 2. Computational details

In this work, Vienna Ab-initio Simulation Package (VASP) [30] was utilized to perform the ground state and the excited state calculations, while the Quantum ATK [31] was used to investigate the electronic transport properties of the GeS monolayer. The Perdew-Burke-Ernzerhof (PBE) of generalized gradient approximation [32] was adopted for the exchange-correlation function. Projector-augmented wave (PAW) pseudopotentials are utilized to describe the electronic wave functions in the core region [33]. The cutoff energy for the plane wave basis expansion was 500 eV. Geometric optimization was performed using the Monkhorst-Pack sampling technique [34] with a special k-point mesh of 32×24×1. The full relaxation of all atoms was allowed until the Hellmann-Feynman force acting on each atom was smaller than 0.01 eV/Å. The single-shot GW (G0W0) approach [35] was employed to calculate the quasi-electronic band structure. We described the screening effects using the plasmon-mode model proposed by Hybertsen and Louie [35]. To ensure the accuracy of our calculations, we have performed convergence tests using various k-mesh, and cutoff energy values for the response functions, as well as the number of empty conduction bands including. Our results (**Figure S1**) demonstrated that the electronic properties are very sensitive to the input parameters, the KPOINTS of 36×27×1, response functions with a cutoff energy of 120 eV, and 120 empty conduction bands were sufficient to achieve convergence for the quasi-particle band gap. Regarding the optical response and the excitonic effects, the dielectric functions were achieved by solving the Bethe-Salpeter equation (BSE) [36] on top of the G0W0 calculations. In this calculation, the 6 highest occupied valence bands (VBs) and 4 lowest unoccupied conduction bands (CBs) are included as a basis for the excitonic states with a photon energy region from 0 eV to 5 eV. In addition, the Lorentz broadening parameter $\gamma$ was set at 40 meV to replace the delta function.

## 3. Results and discussions
## 3.1. Electronic Transport Properties

As a typical benchmark to investigate the impact of strain effects on electronic and optical properties, the geometric structure of pristine monolayer GeS is considered and shown in **Figure 1(a)** and **Table 1**. Similar to the black phosphorene, the monolayer GeS also exhibited a buckled structure with each germanium atom covalently bonded with three adjacent sulfur atoms. The optimized lattice constants are a = 3.665 Å (zigzag direction), and b = 4.471 Å (armchair direction). The calculated parameters are in good agreement with previous theoretical calculations [37-40] and experimental measurements [41].

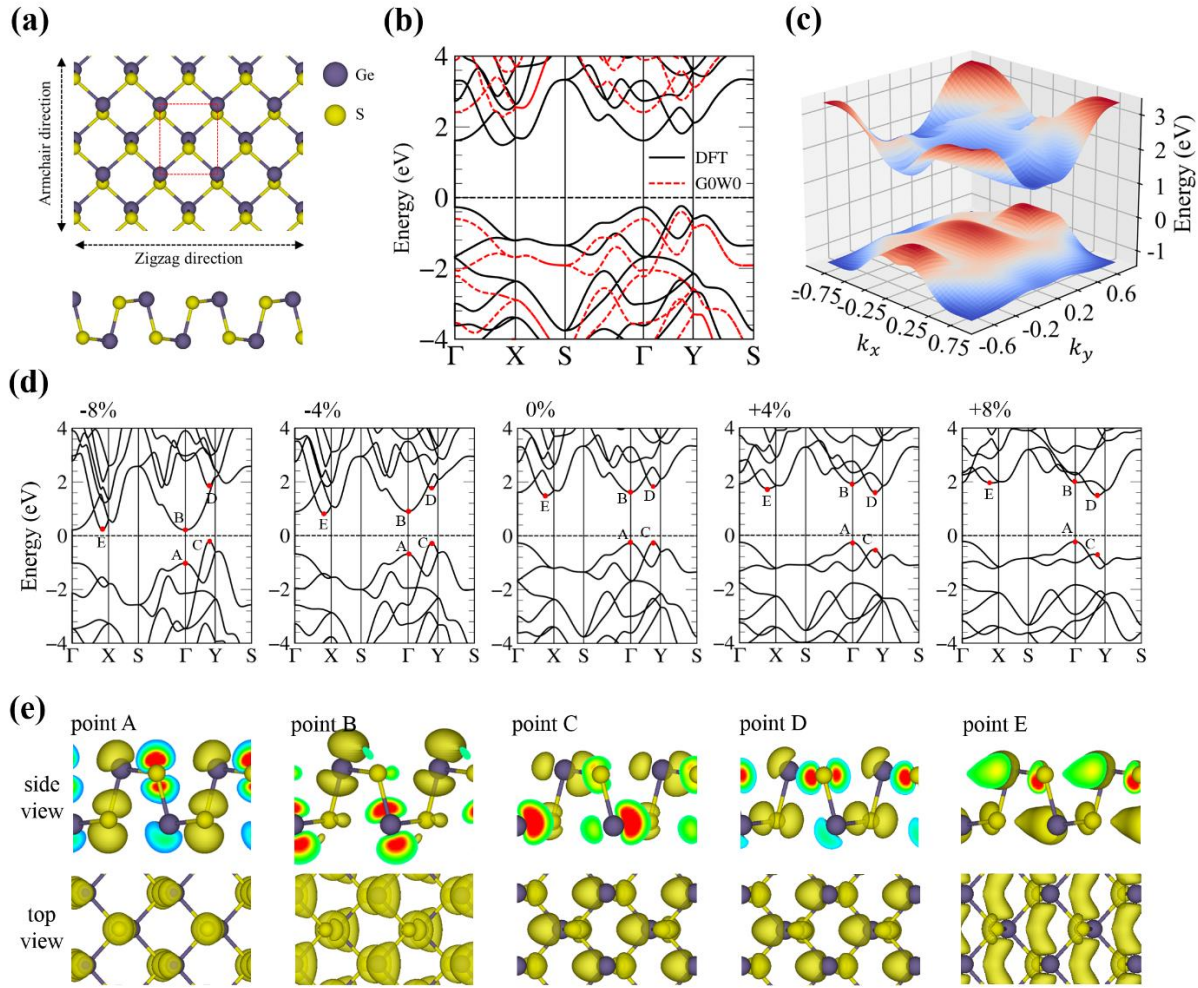

**Figure 1**. (a) Top and side views of geometric structure, (b) the electronic band structure with DFT and G0W0 levels of theory, and (c) the 3D band structure of GeS. (d) The DFT electronic band structure of the GeS monolayer as a function of strain, and (e) the band-decomposed charge density of the corresponding critical points marked by the red dots in (d).

The electronic band structure along the high-symmetry points within DFT and G0W0 levels of theory is shown in **Figure 1(b)**, since the spin-orbit couplings (SOC) only minor influent on the electronic band gap of GeS (**Figure S2**), the relativity effects are ignored in our calculations for the sake of reducing of computational cost. GeS exhibits anisotropic electronic properties, the dispersion of the occupied hole along the $\Gamma - Y$ direction is significant, indicating the small effective mass. The opposite behavior is true for the hole transport along the $\Gamma - X$ direction with relatively flat energy dispersion related to the large effective mass. Similar characteristics are also found for the electron in the unoccupied states. The anisotropic of these bands can be easily detected by the 3D contour plot in **Figure 1(c)**, while the spatial-dependent of electron and hole effective mass of the GeS monolayer, which exhibits the "heart" and "peanut" shapes, are illustrated in

**Figure S3.** GeS monolayer is an indirect band gap of 1.735 eV with the highest occupied state and lowest unoccupied state located between Γ and Y and Γ and X symmetry points, respectively. The electronic band gap is enhanced to 2.665 eV when the electron-electron interactions (GW approximations) are adopted. The theoretical prediction is good in agreement with previous works (**Table 1**).

**Figure 1(d)** depicts the electronic properties of a strained GeS monolayer, with critical points A, B, C, D, and E marked by red dots indicating the band edge states that make the band gap evolution. The corresponding orbital characters for these critical points are shown in **Figures 1(e)** and **Figure S4** and are organized into five categories: (A) out-of-plane interactions of Ge-4$p_z$ and S-3$p_z$ orbitals, (B) out-of-plane hybridizations of Ge-4$p_z$ and S-(3$p_x$, 3$p_y$) orbitals, (C) in-plane couplings Ge-4$p_y$ and S-3s orbitals, (D) interactions of in-plan Ge-4s and S-3$p_y$ orbitals, and (E) in-plane interactions of Ge-4$p_x$ and S-3$p_x$ orbitals. As indicated in **Figure S5** and **Table S1**, compressive strain causes the $d_1$ chemical bonding and h vertical height of Ge-S to increase, significantly reducing the out-of-plane Ge-4$p_z$ and S-3$p_z$, and Ge-4$p_z$ and S-(3$p_x$, 3$p_y$) orbital interactions, leading to a significant shift down of energy levels for the edge states at A and B. On the other hand, the interactions of in-plane couplings of Ge-4$p_y$ and S-3s, and Ge-4s and S-3$p_y$ orbitals increase due to the $d_2$ chemical bonding reduction, resulting in an increase in energy for the edge states C and D. An increase in the α (Ge-S-Ge) angle reduces in-plane interactions of Ge-4$p_x$ and S-3$p_x$ orbitals, causing a downshift in the energy of E critical point. Conversely, the opposite evolution takes place for tensile strains. **Figure S5(b)** displays the evolution of the band gap for the strained GeS monolayer. The electronic band gap decreases linearly with increasing compressive pressure. However, the band gap evolution under tensile strain is more complex. The electronic band gap initially increases to 1.9 eV, accompanied by an indirect-direct transition. However, at higher tensile strains, the gap value dramatically decreases, and the same trend is observed in the band gap evolution with the GW corrections.

To connect the anisotropic band dispersion with the electronic conductance, we further estimated the carrier mobility along the zigzag and armchair directions according to the deformation theory [42, 43]:

$$\mu_{2D} = \frac{e\hbar^3 C_{2D}^i}{k_B T m_i^* m_d E^{i^2}}$$

Here, $m_i^*$ is effective mass along the transport direction, $m_d = \sqrt{m_x^* m_y^*}$ is the average effective mass, $C_{2D}^i$ is the elastic module and can be obtained by the quadratic fitting of the total energy E with respect to the varying of the lattice constant $\Delta l/l_0$ as $(C_{2D}/2)(\Delta l/l_0)^2 = (E - E_0)/S_0$, where $S_0$ is the lattice area at the equilibrium of the 2D lattice. The deformation potential constant $E^i = \partial E_{edge}/\partial \varepsilon$ is obtained by checking the changing of the valence band maximum (VBM) or conduction band minimum (CBM) upon the small lattice compression or lattice expansion along the transport direction. The analytical analysis was established at room temperature T = 300 K. Apparently, the current estimation only depicts the simplest picture of electron-phonon interactions and thus may give overestimated values of the realistic carrier mobility. However, the prediction is accurate enough to capture the anisotropic as well as the tendency of conductance of the system under strains.

Since the effective mass of carriers in the zigzag and armchair directions behaves differently and exhibits two extreme values in these directions (**Figure S3**), we focus on calculating the effective mass and mobility for carriers along the $\Gamma - X$, and $\Gamma - Y$ paths. **Figure 2(a)** illustrates the strain-dependent effective mass of the highest valence hole and lowest conduction electron, which demonstrates a linear decrease under compressive strain. This decrease is reflected in the band curvature shown in **Figure 1(d)** and contributes to enhanced carrier mobility. The carrier effective mass increases significantly under an elongation of the lattice constant, and the anisotropy of carriers along the calculated directions becomes more pronounced. Additionally, the evolution of effective mass is complex. For example, there is a notable jump in $m_h^*$(zigzag) at 2% tensile strain due to the transition of the VBM from $\Gamma - Y$ to $\Gamma$ band edge states, resulting in a shift from light holes to heavy holes.

The calculated carrier mobility of the GeS monolayer at room temperature (T = 300K) according to the compression and elongation of the lattice constant is shown in **Figure 2 (b)**. In the intrinsic case, the relatively small effective mass of electrons and the significant $C_{2D}^i/{E^i}^2$ ratio (**Table S2**) contribute to the high electron carrier mobility of GeS monolayer, with a typical value of $13 \times 10^3$ cm².V⁻¹.s⁻¹ for electrons in the zigzag-direction and about $0.35 \times 10^3$ cm².V⁻¹.s⁻¹ for electrons in the armchair direction. The carrier mobility values for valence holes are lower, with about $0.061 \times 10^3$ cm².V⁻¹.s⁻¹, and $0.036 \times 10^3$ cm².V⁻¹.s⁻¹ for zigzag and armchair directions, respectively. The high electron mobility of 2D GeS is consistent with previous reports [11, 44] and compatible with that of phosphorene [45], but much higher than that of MoS₂ [2], indicating its potential for high-speed electronic applications. The carrier mobility can be efficiently modulated by strain-induced deformations. As expected, carrier mobility decreases upon lattice expansion due to the increasing carrier effective mass and the decreasing of $C_{2D}^i/{E^i}^2$ ratio. Conversely, the mobility of carriers significantly increases with lattice compression. Although the mobility of holes can be controlled via applying external strain, it cannot surpass that of the electrons. Interestingly, under -8% compression, the mobility of electrons of GeS monolayer along a zigzag direction reaches approximately $35 \times 10^3$ cm².V⁻¹.s⁻¹, more than 2.5 times and 400 times larger than the values under free-strain and +8% tensile strain, respectively.

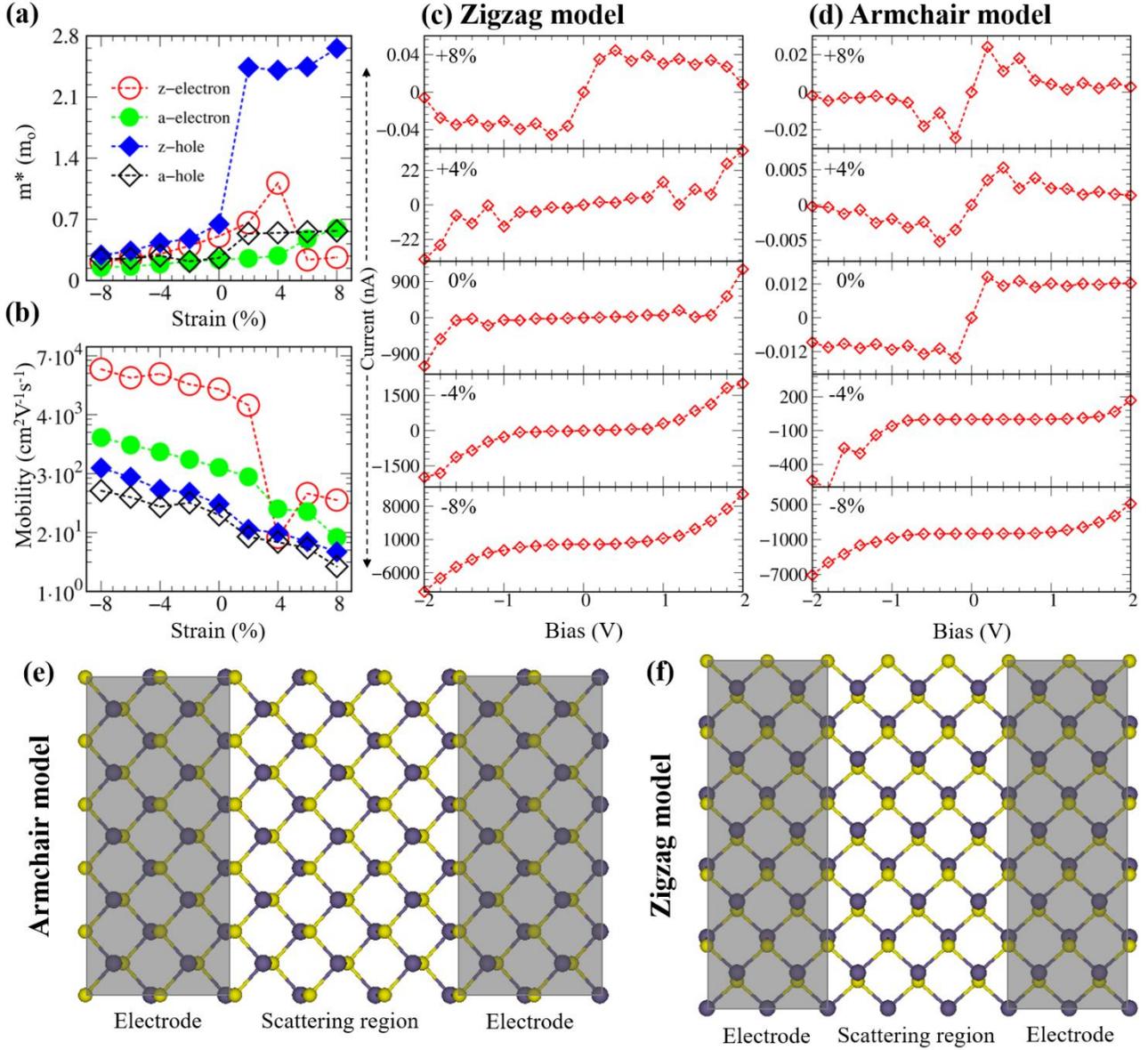

**Figure 2.** (a) The evolutions of the effective mass, and (b) the carrier mobility under biaxial strains. I-V characteristics for (c) zigzag model and (d) armchair model at different biaxial strains. The device model for electron transport along the (e) armchair (armchair model) and (f) zigzag (zigzag model) directions, in which, two- and three-unit cells were used to construct the electrodes, and the active region, respectively.

Due to its exceptionally high carrier mobility, we opted for strained GeS monolayers as the channel material in our device construction. The transport properties have been calculated via the NEGF method as implemented in the Quantum ATK package [31]. 50×50×1 k-points were used for the central region and the electrodes. When a given voltage is applied, the current is allowed to flow across the system. The electric current (I) is further calculated using the Landauer approach [46], and this can be obtained from the integration of the transmission curve as

$$I(V_b) = \frac{2e}{h} \int_{-\infty}^{+\infty} T(E, V_b)[f(E - \mu_L) - f(E - \mu_R)] \, dE,$$

where $f(E - \mu_{L/R})$ is the Fermi-Dirac distribution function of the $L$ and $R$ electrodes, $\mu_{L/R}$ is the chemical potential, which can move up and down according to the Fermi energy, and $T(E, V_b)$ is the transmission function at energy $E$ and bias $V_b$. The expression of $T(E, V_b)$ is as follows:

$$T(E, V_b) = Tr[\Gamma_L(E, V_b) G(E, V_b) \Gamma_R(E, V_b) G^\dagger(E, V_b)],$$

in which, the coupling matrices are given as $\Gamma_{L/R}$, and the retarded and the advanced Green's functions of the scattering region are presented $G^\dagger$ and $G$.

**Figure 2(e)** and **Figure 2(f)** illustrates the fundamental architecture of the GeS device, highlighting its transport characteristics along the zigzag and armchair directions. These transport properties are further depicted in **Figure 2(c)** and **Figure 2(d)**. Given the negligible carrier mobility of holes, our primary focus lies on the transport properties of electrons. To achieve efficient carrier injection and attain optimal device performance, we employ left and right electrodes with an n-type doping concentration of $3 \times 10^{19}$ e/cm². The intrinsic monolayer GeS exhibits a remarkable anisotropic behavior in its transport properties. The I-V curve, when biased along the zigzag direction, resembles that of a characteristic semiconductor, with a peak current of approximately 1200 nA at $V_{bias} = 2$ V. Conversely, carrier transport along the armchair direction is negligible, with the highest current reaching only 0.012 nA at 0.2 V, followed by slight fluctuations at higher applied voltages. These anisotropic transport characteristics of the intrinsic GeS monolayer align well with previous findings [47] and reflect the primary trend in carrier mobility in their respective directions. Both models demonstrate a high sensitivity of the I-V curves to external strain. As the lattice elongation increases, the maximum current intensity of the zigzag and armchair models experiences a sharp decline. Both models exhibit strong negative differential resistance, indicating a diminishing of current intensity with increasing bias voltage, particularly evident under +8% lattice elongation. Interestingly, the I-V curves of the GeS monolayer device under compression consistently exhibit semiconductor characteristics, even for the armchair model. The current intensity of the compressive GeS device experiences a significant enhancement. For the zigzag model, the highest current intensity exceeds 2000 nA and 10000 nA under -4% and -8% compression, respectively, whereas the corresponding values for the armchair model are approximately 170 nA and 5200 nA. These findings indicate that the devices of compressively strained GeS monolayer possess an extremely high sensitivity, making them well-suited for high-speed electronic applications.

**Table 1:** The optimize geometric parameters and the electronic band gap of monolayer Gelium Sulfide. The previous theoretical and experimental measurements are also shown for comparison.

| a(Å) | b(Å) | Fundamental band gap | |
| --- | --- | --- | --- |
| | | DFT | G0W0 |
| 4.471[a] | 3.665[a] | 1.728[a] | 2.661[a] |
| 4.470[b] | 3.666[b] | - | 2.74[b] |
| 4.459[c] | 3.662[c] | 1.90[c] | - |
| 4.33[d] | 3.67[d] | - | - |
| 4.492[e] | 3.62[e] | 1.713[e] | - |
| 4.467[f] | 3.666[f] | 1.722[f] | - |
| 4.474[g] | 3.675[g] | 1.82[g] | - |
| 4.29[h] | 3.64[h] | - | - |

a. The theoretical approach in this work
b. The theoretical approach in Ref [15]
c. The theoretical approach in Ref [48]
d. The theoretical approach in Ref [11]
e. The theoretical approach in Ref [38]
f. The theoretical approach in Ref [39]
g. The theoretical approach in Ref [40]
h. Experimental data for bulk GeS in Ref [41]

## 3.2. Optical Properties and Excitonic Effects

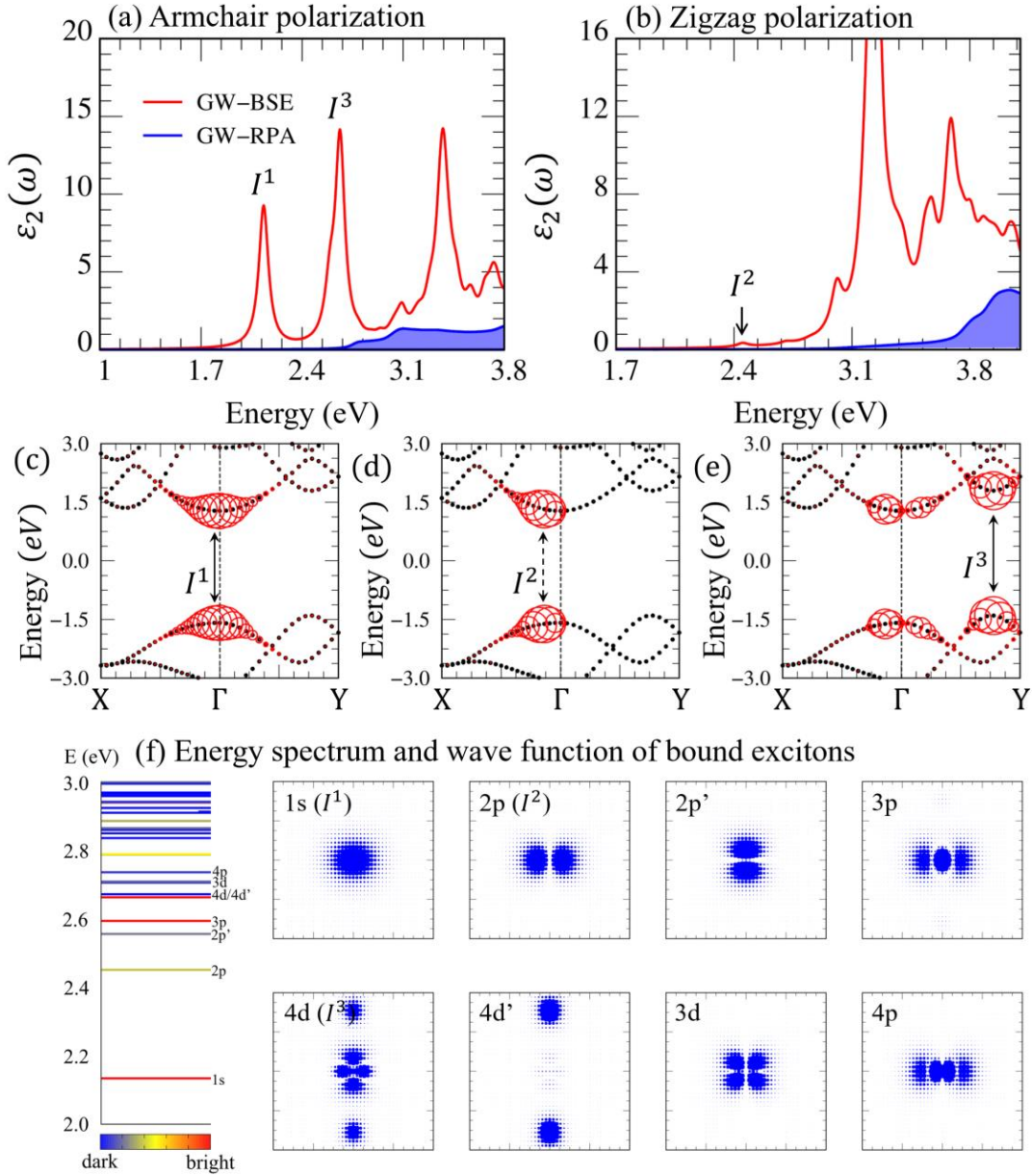

**Figure 3**. The imaginary part of dielectric functions of GeS monolayer along (a) armchair and (b) zigzag polarizations. The red curve indicates the optical spectrum with includes the excitonic effects, while the blue-filled curve excludes these effects. The exciton wave functions are projected onto the electronic band structure for the (c) $I^1$, (d) $I^2$, and (e) $I^3$ excitons, showing the vertical excitation from the VBM to the CBM. The radii of circles represent the contribution of electron–hole pair at that k-point to the *i*th exciton wave function, the dots in the background are the corresponding G0W0 quasi-particle band structures. (f) The exciton energy spectrum of the GeS monolayer and the corresponding k-space distribution of the first eight exciton envelope functions.

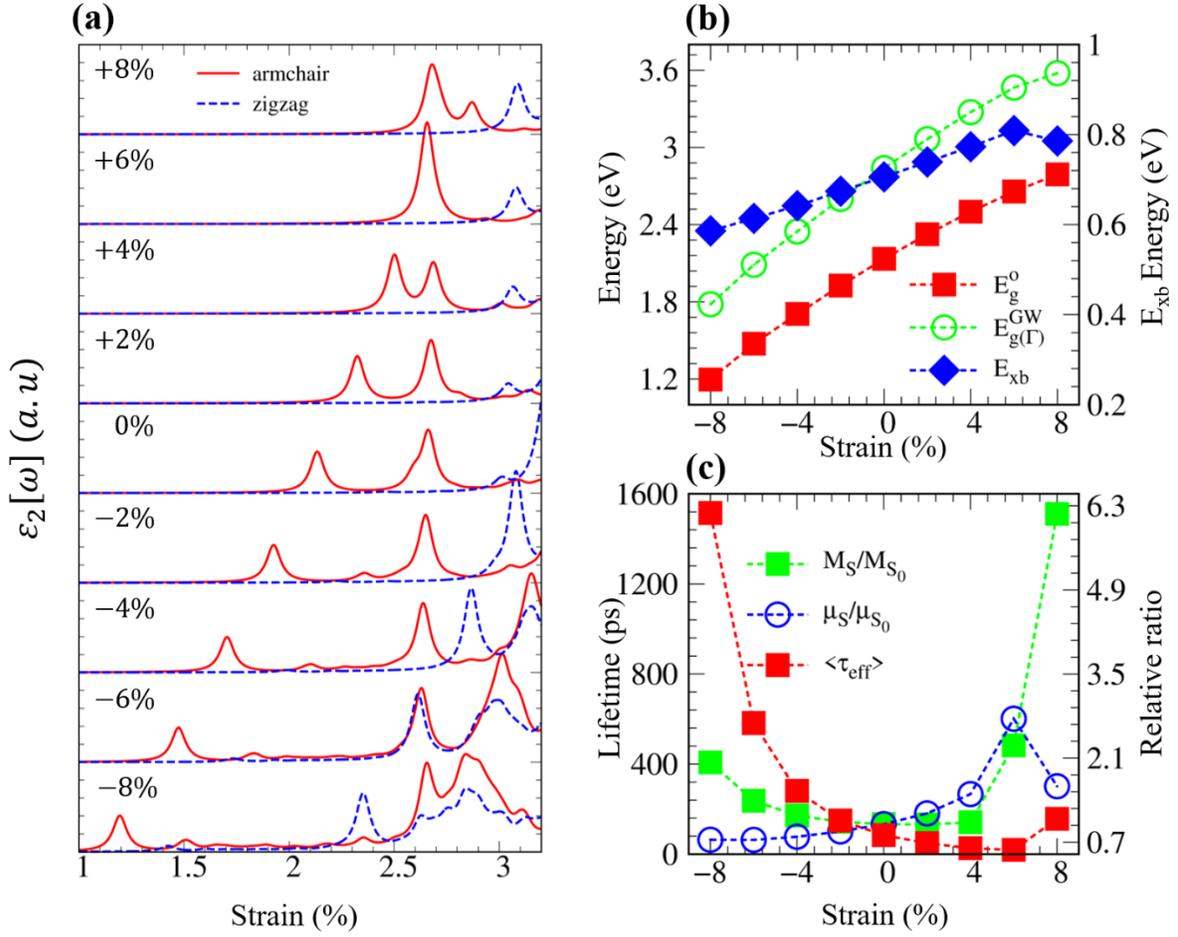

**Figure 4**. (a) The imaginary part of the dielectric function is a function of external strains of the GeS monolayer. (b) The evolution of the optical gap, the direct energy band gap at $\Gamma$ point, and the exciton binding energy under strain. (c) The effective radiative lifetime $\langle \tau_{eff} \rangle$, the oscillation strength ratio $(\mu_S/\mu_{S_0})$ and exciton effective mass ratio $(M_S/M_{S_0})$ of the first bright exciton state as a function of external strain.

**Figure 3** shows the optical properties of the GeS monolayer with and without excitonic effects. The absorbance spectra exhibit significant anisotropy due to the non-uniform environment along the armchair and zigzag directions. The low-frequency optical properties are primarily influenced by the armchair-polarization. In the absence of strain, the optical properties of GeS along the armchair and zigzag directions are characterized by three exciton states, denoted as $I^1$, $I^3$, and $I^2$ in **Figures 3(a)** and **3(b)**, respectively. These excitonic states, correspondingly, originated from the coupling of excited holes and electrons at the $\Gamma$ valley $(S_{3p_z} \to Ge_{4p_z})$, the band edge state at $\Gamma - Y$ path $(S_{3p_{xy}} \to Ge_{4p_{xy}})$, and the critical point along the $\Gamma - X$ direction $(Ge_{4p_z} \to S_{3p_{xy}})$ as indicated by the fat band in **Figures 3(c)** to **3(e)**. The $I^2$ exciton state is rather weak since the transition from the occupied state to the unoccupied one of this exciton is approximately forbidden, which arises from the different parity of its excited hole and excited electron [49], the opposite behavior is true for $I^1$ and $I^3$ excitons. To assess the strength of excitonic effects, we calculated exciton binding energy as the energy difference between the optical gap and the fundamental direct GW band gap at Γ. The exciton binding energy of the GeS monolayer is about 0.706 eV (**Table S4**), and is consistent with previous reports [13]. The large binding energies and the significant modifications in the adsorption spectra (compared with GW-RPA spectra) indicate that these exciton states are strong and potentially stable at high temperatures. The dissociation temperature $T_d$ ($T_d \approx 0.1 E_b/K_B$) for the $I^1$ exciton is around 820K, which is higher than room temperature.

To better understand the character of specific exciton states, **Figure 3(f)** illustrates the energy diagram of the bound exciton states in the GeS monolayer, as well as the k-space distribution of the squared amplitude of the free electron-hole pairs that constitute the exciton wave functions in the Brillouin zone. In addition to the bright exciton states, the peculiar parity symmetries of band edges result in the lowest optical transition dipole being forbidden, leading to numerous dark states that are visible as blue-grey colors in **Figure 3(f)**. Although not detectable in the optical absorbance spectra, the dark exciton states are important as they provide fingerprints for the optical properties of typical materials. The nodal structures of these excitonic wave functions reveal a hydrogen-like series of states with clear angular momentum assignments. Interestingly, the excitonic energy diagram does not follow the Rydberg series for the 2D hydrogenic model, in which excitons with higher azimuthal quantum numbers have lower energies than those with smaller azimuthal quantum numbers. For instance, the energy of 4d exciton is smaller than that of the 3d and 4p ones, this behavior is universal in 2D materials due to their unique screening [50]. Another noteworthy feature is the degeneracy of the 2p states resulting from the in-plane anisotropy, which is a characteristic of the GeS monolayer and other 2D materials [51, 52].

**Figure 4 (a)** illustrates the optical excitation of GeS when subjected to biaxial strain, while **Figure 4(b)** summarizes the changes in the optical gap and the direct GW electronic band gap at Γ, as well as the exciton binding energy. For the sake of simplification, we only focus here on the features of the first bright exciton $I^1$. When compressed, the optical gap decreases due to the reduction of the direct electronic band gap at the Γ point. The exciton binding energy of prominent excitations and their intensity exhibits a significant alteration. The most notable feature is that the exciton binding energy of GeS under compressive strain is weaker than that of the strain-free and lattice-expanded cases. This is primarily due to the enhanced screening ability (increase in static dielectric constant $\varepsilon(0)$ as shown in **Table S3**) or the reduction of the electronic band gap. Additionally, the anisotropy of the optical absorbance spectrum along the armchair and zigzag polarizations at P = -6% gradually decreases compared to that of the intrinsic case, but they start to distinguish at the higher strain applied. This is due to the evolution of the electronic anisotropy of the GeS monolayer under compression, as indicated by the contour plots of the direct valence to conduction band transitions in **Figure 5**.

Conversely, when subjected to lattice expansion, the optical gap increases, and the exciton binding energy and the anisotropy of the optical spectrum exhibit opposite changes to those observed under compression. The changes in optical properties of the GeS monolayer upon elongation are rather complicated, as the optical gap gradually increases but then slowly decreases at higher applied strain. The distinguishing of the optical properties along the armchair and zigzag directions are more obvious due to the anisotropic electronic wave functions of tensile strained GeS monolayer (**Figure 5**). Moreover, the exciton binding energy and the intensity of the first bright state I$^1$ significantly increase. The evolution of binding energy of the first bright state could be deduced by the increase of the electronic band gap or decrease of the dielectric screening environment (**Table S3**). On the other hand, the enhancement of its intensity can be explained as follows: the excited hole mostly localized around the Phosphorus atom, while the excited electron relied around the sulfur atom of the opposite plane as shown in the two-first panels of **Figure 1(d),** indicated in the reciprocal space in **Figure 3(c)** and be replotted in **Diagram S6.** The significant reduction of monolayer thickness of GeS upon lattice expansion induces the hole and electron to come closer, thereby enhancing the electron-hole overlap, and the transition probabilities, as well as exciton binding energy. However, the impact of electron-hole physical distance does not always express the linear relation, the binding energy of excitons and their oscillation strength begin to decrease when the critical strain reaches +8%. This phenomenon occurs because the nature of the first bright state gradually shifts from the 1s state to the 5d state as the GeS monolayer undergoes elongation. This information is illustrated in **Figure 5**.

**Table 2.** The effective excition lifetime calculate at room temperature ($\tau_{eff}^{RT}$) of intrinsic GeS monolayer. The previous calculation and experimental values are also listed for comparison.

| Materials | $\tau_{eff}^{RT}$ |
|---|---|
| GeS | 84.86 (ps)[a] |
| Phosphorene | 179.05 (ps)[a] |
| | 194.21 (ps)[b] |
| | 221.35 (ps)[c] |
| MoS2 | 0.83 (ns)[a] |
| | 0.82 (ns)[d] |
| | 0.85 (ns)[e] |
| MoSe2 | 0.87 (ns)[a] |
| | 0.80 (ns)[d] |
| | 0.90 (ns)[f] |

a. Current theoretical prediction using the Fermi-golden rule
b. Theoretical prediction using Hefei-NAMD code in Ref [53]
c. Experimental measurement in Ref [54]
d. Theoretical prediction using the Fermi-golden rule in Ref [55]
e. Experimental measurement from Ref [56]
f. Experimental measurement from Ref [57]

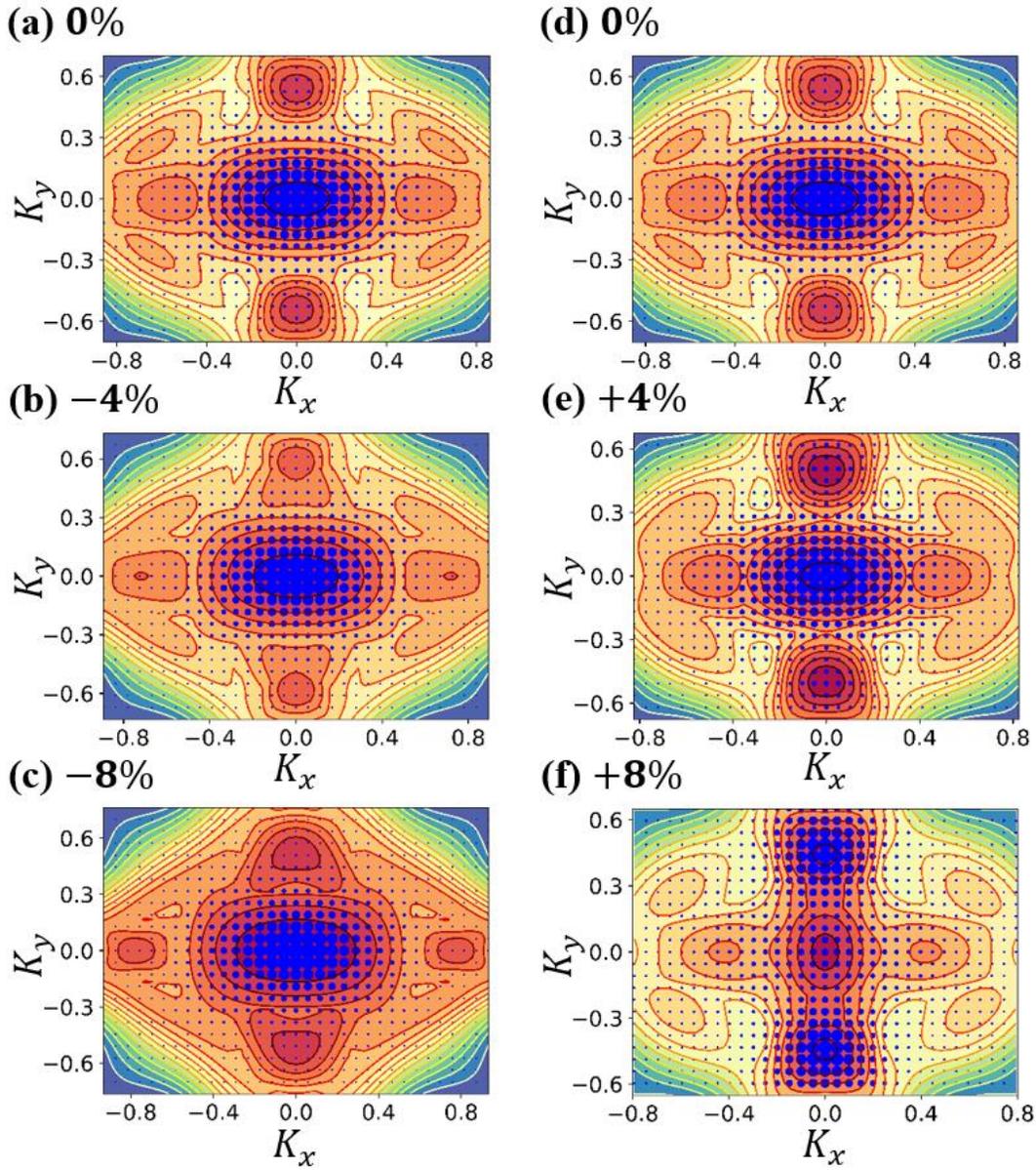

**Figure 5**. The direct valence to conduction band transition energies in the first Briulouin zone of strained-GeS monolayer is plotted in the color map. The red area indicated the low energy regime, while the green one illustrated the high-energy one. The wave functions of the first exciton state are visualized as blue circles. As the lattice constant is compressed, the electronic functions become more isotropic, and the exciton state I[1] corresponds to a 1s orbital. Conversely, with an elongated lattice constant, the wave functions along the zigzag and armchair directions exhibit noticeable differences. The energies of 1s and 3d become closer to each other and the latter will surpass the former at +8% elongation of the lattice constant, resulting in switching their nature.

Based on the above discussion, it was found that strain plays a vital role in modifying the probability of recombination for excited states. Specifically, we conducted additional analysis on the radiative lifetimes of excitons in monolayer GeS. It is worth noting that the short lifetime of excitons can be advantageous for internal quantum efficiency and telecommunications applications. Conversely, the ultra-long timescale of electron-hole recombination is highly beneficial for advanced optoelectronic and thin-film photovoltaic cells. Using the methodology developed for assessing radiative exciton lifetimes in two-dimensional materials [58], the radiative lifetime $\langle \tau_S \rangle$ at room temperature (T = 300 K) of exciton states S is defined as follows:

$$\langle \tau_S \rangle = \left( \frac{8\pi e^2 E_S(0)}{\hbar^2 c} \frac{\mu_S^2}{A_{uc}} \right)^{-1} \frac{3}{4} \left( \frac{E_S(0)^2}{2M_S c^2} \right)^{-1} k_B T,$$

where $A_{uc}$ is the area of the unit cell, $\mu_S^2$ is the square modulus of the BSE exciton transition dipole divided by the number of 2D k-points, and $E_S(0)$ is the exciton energy calculated using the BSE method, and $M_S = m_e^* + m_h^*$ is the effective mass of exciton. It is important to note that $m_{e(h)}^*$ here indicated the effective mass of the excited electron (hole) related to the exciton bound-state, but not for the effective mass of the CBM (VBM) as discussed in the electronic transport section. At zero strain, the exciton lifetime of $I^1$, $I^2$, and $I^3$ exciton is about 84.86 ps, 137.63 ns, and 21.17 ps. The ultra-long lifetime of dark $I^2$ exciton arises from its extremely small dipole strength.

With the assume the presence of perfect thermalization of the exciton states, we further define an effective radiative lifetime $\langle \tau_{eff} \rangle$ by averaging the decay rates over the lowest energy bright and dark excitons:

$$\langle \tau_{eff} \rangle^{-1} = \frac{\sum_S \langle \tau_S \rangle^{-1} e^{-E_S(0)/k_B T}}{\sum_S e^{-E_S(0)/k_B T}}.$$

The calculated effective exciton lifetime of the intrinsic GeS monolayer is shown in **Table 2**. We also add the results of other 2D materials using the same and different schemes of the current and previous works for comparison. The overall agreement of this approach and other theoretical predictions and experimental measurements is quite good, and thus, our calculations are reliable. The GeS monolayer exhibits an effective exciton lifetime of approximately 84.86 ps, which is close to that of $I_1$ and comparable to blue phosphorene (179.05 ps), but faster than $MoS_2$ (0.83 ns) and $MoSe_2$ (0.87 ns). The dependence of the effective exciton lifetime on strain is shown in **Figure 4(c)**, we also include the evolution of the effective mass and relative oscillation strength of the bright exciton state $I^1$ for comparison since it contributes significantly to the effective exciton lifetime. Generally, the compressive strain leads to a rapid increase in $\langle \tau_{eff} \rangle$ due to decreased oscillation strength and increased effective mass. For example, at 0% strain, $\langle \tau_{eff} \rangle$ is approximately 84.86 ps, while at -4% and -8% strains, the $\langle \tau_{eff} \rangle$ values are approximately 280.52 ps and 1514.59 ps, respectively. In contrast, even though a dramatic enhancement in effective exciton mass, the radiative lifetime only gradually decreases with tensile strain applied and starts to enhance at +8% strain, which is most likely due to a slightly changing in electron-hole recombination intensity. The smallest effective radiative lifetime is about 19 ps at +6% tensile strain, which is 4.5 times and 80 times faster than that of intrinsic and -8% compressive strain.

## 4. Conclusions

To summarize, our study focused on examining the impact of biaxial strain on the electronic transport properties and exciton radiative lifetime of the GeS monolayer. The theoretical works are based on the delicate combination of high-precise simulations and the appropriate theoretical models, such as DFT, MBPT, NEGF, the deformation-potential theory for carrier mobility, and the Fermi-golden rule for exciton lifetime. Our findings revealed a significant enhancement in the I-V characteristic when the lattice is compressed due to an improvement in carrier mobility. The optical gap, the anisotropic optical properties, the absorption coefficient, and the excition binding energy are strongly dependent on the applied biaxial strains. Moreover, the strain also can finely adjust the time scale of electron-hole recombination. With the electronic and optical properties that can be flexibly modified via strain engineering, the GeS monolayer may hold great potential for high-tech applications such as ultrafast FETs, PVs, and optoelectronic applications.

# Acknowledgments

This research was funded by the Vietnam Ministry of Education and Training under grant number B2023-TCT-03.

# References

[1] A. K. Geim and K. S. Novoselov, "The rise of graphene," *Nature materials,* vol. 6, pp. 183-191, 2007.

[2] B. Radisavljevic, A. Radenovic, J. Brivio, V. Giacometti, and A. Kis, "Single-layer MoS2 transistors," *Nature nanotechnology,* vol. 6, pp. 147-150, 2011.

[3] X. Li, J. Dong, J. C. Idrobo, A. A. Puretzky, C. M. Rouleau, D. B. Geohegan*, et al.*, "Edge-controlled growth and etching of two-dimensional GaSe monolayers," *Journal of the American Chemical Society,* vol. 139, pp. 482-491, 2017.

[4] T. Cao, Z. Li, and S. G. Louie, "Tunable magnetism and half-metallicity in hole-doped monolayer GaSe," *Physical review letters,* vol. 114, p. 236602, 2015.

[5] J. D. Caldwell, I. Aharonovich, G. Cassabois, J. H. Edgar, B. Gil, and D. Basov, "Photonics with hexagonal boron nitride," *Nature Reviews Materials,* vol. 4, pp. 552-567, 2019.

[6] K. Zhang, Y. Feng, F. Wang, Z. Yang, and J. Wang, "Two dimensional hexagonal boron nitride (2D-hBN): synthesis, properties and applications," *Journal of Materials Chemistry C,* vol. 5, pp. 11992-12022, 2017.

[7] A. Carvalho, M. Wang, X. Zhu, A. S. Rodin, H. Su, and A. H. Castro Neto, "Phosphorene: from theory to applications," *Nature Reviews Materials,* vol. 1, pp. 1-16, 2016.

[8] D. Tan, H. E. Lim, F. Wang, N. B. Mohamed, S. Mouri, W. Zhang*, et al.*, "Anisotropic optical and electronic properties of two-dimensional layered germanium sulfide," *Nano Research,* vol. 10, pp. 546-555, 2017.

[9] E. Sutter, B. Zhang, M. Sun, and P. Sutter, "Few-Layer to Multilayer Germanium(II) Sulfide: Synthesis, Structure, Stability, and Optoelectronics," *ACS Nano,* vol. 13, pp. 9352-9362, 2019/08/27 2019.
[10] D. Tan, H. E. Lim, F. Wang, N. B. Mohamed, S. Mouri, W. Zhang*, et al.*, "Anisotropic optical and electronic properties of two-dimensional layered germanium sulfide," *Nano Research,* vol. 10, pp. 546-555, 2017/02/01 2017.
[11] F. Li, X. Liu, Y. Wang, and Y. Li, "Germanium monosulfide monolayer: a novel two-dimensional semiconductor with a high carrier mobility," *Journal of Materials Chemistry C,* vol. 4, pp. 2155-2159, 2016.
[12] S. Yu, H. D. Xiong, K. Eshun, H. Yuan, and Q. Li, "Phase transition, effective mass and carrier mobility of MoS2 monolayer under tensile strain," *Applied Surface Science,* vol. 325, pp. 27-32, 2015/01/15/ 2015.
[13] L. C. Gomes, P. Trevisanutto, A. Carvalho, A. Rodin, and A. C. Neto, "Strongly bound Mott-Wannier excitons in GeS and GeSe monolayers," *Physical Review B,* vol. 94, p. 155428, 2016.
[14] Y.-H. Chan, D. Y. Qiu, F. H. da Jornada, and S. G. Louie, "Giant exciton-enhanced shift currents and direct current conduction with subbandgap photo excitations produced by many-electron interactions," *Proceedings of the National Academy of Sciences,* vol. 118, p. e1906938118, 2021.
[15] L. Xu, M. Yang, S. J. Wang, and Y. P. Feng, "Electronic and optical properties of the monolayer group-IV monochalcogenides M X (M= Ge, Sn; X= S, Se, Te)," *Physical Review B,* vol. 95, p. 235434, 2017.
[16] V. K. Dien, O. K. Le, V. Chihaia, M.-P. Pham-Ho, and D. N. Son, "Monolayer transition-metal dichalcogenides with polyethyleneimine adsorption," *Journal of Computational Electronics,* vol. 20, pp. 135-150, 2021.
[17] V. K. Dien, N. T. Han, W. Bang-Li, K. I. Lin, and M. F. Lin, "Tuning of the Electronic and Optical Properties of Monolayer GaSe Via Strain," *Advanced Theory and Simulations,* p. 2200950, 2023.
[18] N. T. T. Tran, D. K. Nguyen, O. E. Glukhova, and M.-F. Lin, "Coverage-dependent essential properties of halogenated graphene: A DFT study," *Scientific reports,* vol. 7, p. 17858, 2017.
[19] V. K. Dien, W.-B. Li, K.-I. Lin, N. T. Han, and M.-F. Lin, "Electronic and optical properties of graphene, silicene, germanene, and their semi-hydrogenated systems," *RSC advances,* vol. 12, pp. 34851-34865, 2022.
[20] H.-C. Chung, C.-P. Chang, C.-Y. Lin, and M.-F. Lin, "Electronic and optical properties of graphene nanoribbons in external fields," *Physical Chemistry Chemical Physics,* vol. 18, pp. 7573-7616, 2016.
[21] W. Zhou, X. Zou, S. Najmaei, Z. Liu, Y. Shi, J. Kong*, et al.*, "Intrinsic structural defects in monolayer molybdenum disulfide," *Nano letters,* vol. 13, pp. 2615-2622, 2013.
[22] P. K. Chow, R. B. Jacobs-Gedrim, J. Gao, T.-M. Lu, B. Yu, H. Terrones*, et al.*, "Defect-Induced Photoluminescence in Monolayer Semiconducting Transition Metal Dichalcogenides," *ACS Nano,* vol. 9, pp. 1520-1527, 2015/02/24 2015.
[23] V. Van On, P. T. Bich Thao, L. N. Thanh, and N. T. Tien, "Insights on modulating electronic and transport properties of the sawtooth–sawtooth penta-SiC2 nanoribbons under uniaxial small strain by first-principles calculations," *AIP Advances,* vol. 12, 2022.
[24] Y. Li, T. Wang, M. Wu, T. Cao, Y. Chen, R. Sankar*, et al.*, "Ultrasensitive tunability of the direct bandgap of 2D InSe flakes via strain engineering," *2D Materials,* vol. 5, p. 021002, 2018.
[25] J.-H. Wong, B.-R. Wu, and M.-F. Lin, "Strain effect on the electronic properties of single layer and bilayer graphene," *The Journal of Physical Chemistry C,* vol. 116, pp. 8271-8277, 2012.
[26] I. M. Datye, A. Daus, R. W. Grady, K. Brenner, S. Vaziri, and E. Pop, "Strain-Enhanced Mobility of Monolayer MoS2," *Nano Letters,* vol. 22, pp. 8052-8059, 2022/10/26 2022.

[27]　R. G. Parr, "Density functional theory," *Annual Review of Physical Chemistry,* vol. 34, pp. 631-656, 1983.

[28]　D. Wing, J. B. Haber, R. Noff, B. Barker, D. A. Egger, A. Ramasubramaniam*, et al.*, "Comparing time-dependent density functional theory with many-body perturbation theory for semiconductors: Screened range-separated hybrids and the G W plus Bethe-Salpeter approach," *Physical Review Materials,* vol. 3, p. 064603, 2019.

[29]　A. Pecchia, G. Penazzi, L. Salvucci, and A. Di Carlo, "Non-equilibrium Green's functions in density functional tight binding: method and applications," *New Journal of Physics,* vol. 10, p. 065022, 2008.

[30]　J. Hafner, "Ab-initio simulations of materials using VASP: Density-functional theory and beyond," *Journal of Computational Chemistry,* vol. 29, pp. 2044-2078, 2008.

[31]　S. Smidstrup, T. Markussen, P. Vancraeyveld, J. Wellendorff, J. Schneider, T. Gunst*, et al.*, "QuantumATK: an integrated platform of electronic and atomic-scale modelling tools," *Journal of Physics: Condensed Matter,* vol. 32, p. 015901, 2019.

[32]　J. P. Perdew, K. Burke, and M. Ernzerhof, "Generalized Gradient Approximation Made Simple," *Physical Review Letters,* vol. 77, pp. 3865-3868, 10/28/ 1996.

[33]　G. Kresse and D. Joubert, "From ultrasoft pseudopotentials to the projector augmented-wave method," *Physical Review B,* vol. 59, pp. 1758-1775, 01/15/ 1999.

[34]　P. Wisesa, K. A. McGill, and T. Mueller, "Efficient generation of generalized Monkhorst-Pack grids through the use of informatics," *Physical Review B,* vol. 93, p. 155109, 04/06/ 2016.

[35]　M. S. Hybertsen and S. G. Louie, "Electron correlation in semiconductors and insulators: Band gaps and quasiparticle energies," *Physical Review B,* vol. 34, p. 5390, 1986.

[36]　M. Rohlfing and S. G. Louie, "Electron-hole excitations and optical spectra from first principles," *Physical Review B,* vol. 62, pp. 4927-4944, 08/15/ 2000.

[37]　N. T. Kaner, Y. Wei, Y. Jiang, W. Li, X. Xu, K. Pang*, et al.*, "Enhanced shift currents in monolayer 2D GeS and SnS by strain-induced band gap engineering," *ACS omega,* vol. 5, pp. 17207-17214, 2020.

[38]　Y. Xu, K. Xu, and H. Zhang, "First-principles calculations of angular and strain dependence on effective masses of two-dimensional phosphorene analogues (monolayer α-phase group-IV monochalcogenides MX)," *Molecules,* vol. 24, p. 639, 2019.

[39]　T. Hu and J. Dong, "Two new phases of monolayer group-IV monochalcogenides and their piezoelectric properties," *Physical Chemistry Chemical Physics,* vol. 18, pp. 32514-32520, 2016.

[40]　K. D. Pham, C. V. Nguyen, H. V. Phuc, T. V. Vu, N. V. Hieu, B. D. Hoi*, et al.*, "Ab-initio study of electronic and optical properties of biaxially deformed single-layer GeS," *Superlattices and Microstructures,* vol. 120, pp. 501-507, 2018.

[41]　W. Zachariasen, "The crystal lattice of germano sulphide, GeS," *Physical Review,* vol. 40, p. 917, 1932.

[42]　J. Bardeen and W. Shockley, "Deformation potentials and mobilities in non-polar crystals," *Physical review,* vol. 80, p. 72, 1950.

[43]　S. Bruzzone and G. Fiori, "Ab-initio simulations of deformation potentials and electron mobility in chemically modified graphene and two-dimensional hexagonal boron-nitride," *Applied Physics Letters,* vol. 99, p. 222108, 2011.

[44]　M. Cheng and J. Guan, "Two-dimensional Haeckelite GeS with high carrier mobility and exotic polarization orders," *Physical Review Materials,* vol. 5, p. 054005, 05/14/ 2021.

[45]　R. Fei and L. Yang, "Strain-Engineering the Anisotropic Electrical Conductance of Few-Layer Black Phosphorus," *Nano Letters,* vol. 14, pp. 2884-2889, 2014/05/14 2014.

[46]　E. G. Emberly and G. Kirczenow, "Landauer theory, inelastic scattering, and electron transport in molecular wires," *Physical Review B,* vol. 61, p. 5740, 2000.

[47] Y. Ding, Y.-S. Liu, G. Yang, Y. Gu, Q. Fan, N. Lu, *et al.*, "High-Performance Ballistic Quantum Transport of Sub-10 nm Monolayer GeS Field-Effect Transistors," *ACS Applied Electronic Materials,* vol. 3, pp. 1151-1161, 2021/03/23 2021.

[48] N. T. Kaner, Y. Wei, Y. Jiang, W. Li, X. Xu, K. Pang, *et al.*, "Enhanced Shift Currents in Monolayer 2D GeS and SnS by Strain-Induced Band Gap Engineering," *ACS Omega,* vol. 5, pp. 17207-17214, 2020/07/21 2020.

[49] G. Antonius, D. Y. Qiu, and S. G. Louie, "Orbital symmetry and the optical response of single-layer MX monochalcogenides," *Nano letters,* vol. 18, pp. 1925-1929, 2018.

[50] D. Y. Qiu, F. H. Da Jornada, and S. G. Louie, "Screening and many-body effects in two-dimensional crystals: Monolayer $MoS_2$," *Physical Review B,* vol. 93, p. 235435, 2016.

[51] D. Y. Qiu, F. H. Da Jornada, and S. G. Louie, "Optical spectrum of $MoS_2$: many-body effects and diversity of exciton states," *Physical review letters,* vol. 111, p. 216805, 2013.

[52] L. Yang, J. Deslippe, C.-H. Park, M. L. Cohen, and S. G. Louie, "Excitonic effects on the optical response of graphene and bilayer graphene," *Physical review letters,* vol. 103, p. 186802, 2009.

[53] X. Wang, W. Gao, and J. Zhao, "Strain modulation of the exciton anisotropy and carrier lifetime in black phosphorene," *Physical Chemistry Chemical Physics,* vol. 24, pp. 10860-10868, 2022.

[54] J. Yang, R. Xu, J. Pei, Y. W. Myint, F. Wang, Z. Wang, *et al.*, "Optical tuning of exciton and trion emissions in monolayer phosphorene," *Light: Science & Applications,* vol. 4, pp. e312-e312, 2015.

[55] M. Palummo, M. Bernardi, and J. C. Grossman, "Exciton Radiative Lifetimes in Two-Dimensional Transition Metal Dichalcogenides," *Nano Letters,* vol. 15, pp. 2794-2800, 2015/05/13 2015.

[56] H. Shi, R. Yan, S. Bertolazzi, J. Brivio, B. Gao, A. Kis, *et al.*, "ACS Nano 2013, 7, 1072–1080 DOI: 10.1021/nn303973r [ACS Full Text ACS Full Text]," *Google Scholar There is no corresponding record for this reference*.

[57] N. Kumar, Q. Cui, F. Ceballos, D. He, Y. Wang, and H. Zhao, "Exciton-exciton annihilation in $MoSe_2$ monolayers," *Physical Review B,* vol. 89, p. 125427, 2014.

[58] C. Robert, D. Lagarde, F. Cadiz, G. Wang, B. Lassagne, T. Amand, *et al.*, "Exciton radiative lifetime in transition metal dichalcogenide monolayers," *Physical review B,* vol. 93, p. 205423, 2016.